\documentclass[showpacs,preprintnumbers,amsmath,amssymb,prl,twocolumn]{revtex4}

\usepackage{graphicx}
\usepackage{dcolumn}
\usepackage{bm}
\def\comment#1{}

\begin{document}

\title{Deconfinement mechanism in three dimensions 
for gauge fields coupled to bosonic matter fields with fundamental 
charge}
\author{Flavio S. Nogueira}
\email{nogueira@physik.fu-berlin.de}
\affiliation{Institut f\"ur Theoretische Physik,
Freie Universit\"at Berlin, Arnimallee 14, D-14195 Berlin, Germany}
\author{Asle Sudb{\o}}
\email{asle.sudbo@phys.ntnu.no}
\affiliation{Department of Physics, Norwegian University of
Science and Technology, N-7491 Trondheim, Norway}
\date{Received \today}

\begin{abstract}
We propose a mechanism by which electric charges deconfine 
in an Abelian Higgs model with matter fields belonging to 
the fundamental representation of the gauge group. 
Kosterlitz-Thouless like recursion relations for a
scale-dependent stiffness parameter and fugacity are given,
showing that for a logarithmic potential between point 
charges in any dimension, there exists a stable fixed point 
at zero fugacity, with a dimensionality dependent universal 
jump in the stiffness parameter at the phase transition. 
\end{abstract}

\pacs{11.15.-q,11.15.Ha,11.15.Tk}
\maketitle


It is well known that a pure Maxwell theory with gauge fields arising 
from a {\it compact} $U(1)$ gauge group permanently confines 
electric charges in three dimensions \cite{Polyakov}. In such a case,
specifically in the absence of matter fields coupled to the gauge fields,  
the Wilson loop is a good probe for confinement and satisfies 
the area law, with a linearly confining potential between electric 
test charges. The periodic character of the gauge fields induces 
topological defects in the theory. In the case of compact Maxwell theory 
in three dimensions these defects are magnetic monopoles. This is easily 
seen as the $2\pi$ jumps can be accounted for by writing the field strength 
as

\begin{equation}
\label{F}
F_{\mu\nu}=f_{\mu\nu}-2\pi\epsilon_{\mu\nu\lambda}
\partial_\lambda\int d^3y D(x-y)n(y),
\end{equation}
where $f_{\mu\nu}$ is the non-singular part of the field strength, 
$n(x)=\sum_i q_i\delta^3(x-x_i)$ is the magnetic charge density and 

\begin{equation}
-\partial^2 D(x-y)=\delta^3(x-y).
\end{equation}
Thus, the dual field strength 
$F^*_\mu=\epsilon_{\mu\nu\lambda}F_{\nu\lambda}/2$ satisfies 

\begin{equation}
\partial_\mu F^*_\mu(x)=2\pi n(x).
\end{equation}

In the absence of monopoles the potential between electric test 
charges is given simply by the Coulomb interaction in two 
space dimensions:

\begin{equation}
\label{pot1}
V(R)\sim\ln R. 
\end{equation}
The monopoles of the dual theory, on the other hand, interact through a 
Coulomb potential in {\it three dimensions}, i.e., $V_{\rm mon}\sim 1/r$. 
In order to show that in the compact Maxwell theory electric charges 
never deconfine, we have to study a classical Coulomb gas of 
monopoles in three dimensions. Polyakov \cite{Polyakov} carried out 
this study by showing that for low fugacity the Coulomb gas maps 
onto a sine-Gordon theory, which he solved in a saddle point 
approximation in the presence of external sources corresponding 
to the test electric charges. The end result is a potential 
between electric test charges $V(R) \sim R$ in two space dimensions. 
Thus, the monopoles act in such a way as to produce an anti-screening 
effect in the potential between electric charges. The force between 
the electric charges, which decreases as $1/R$ in two space dimensions 
for the non-compact Maxwell theory, does not vary with the distance 
any longer in the case of compact Maxwell theory. 

Note that in two space-dimensions, the absence of monopoles in the 
gauge-field produces a potential between test charges which is given 
by $V(R) \sim \ln R$.  In the presence of fluctuating matter fields 
in three space-time dimensions (two space dimensions), this  is 
changed to $V(R) \sim 1/R$, due to the presence of an anomalous 
scaling dimension of the gauge field. The gauge-field propagator in 
$d$ space-time dimensions in the presence of critical matter-field 
fluctuations is given by $D(|x|) \sim 1/|x|^{d-2+\eta_A}$, where 
$\eta_A$ is the anomalous scaling dimension of the gauge-field,
gauge-invariance dictates that $\eta_A=4-d$ \cite{Herbut}, and 
$d$ is the space-time dimensionality. Therefore, we see that 
$D(|x|) \sim 1/|x|^2$ when $d \in (2,4]$. This is the same as we would 
get with or without matter-field fluctuations in four-dimensional 
space-time. 
Hence, we see that the role of the anomalous scaling of the gauge 
field is to essentially produce extra space dimensions. It is as if the  
test charges were living in three space dimensions, but confined to 
moving in two.
 
An alternative derivation of Polyakov's result amounts to showing that a 
Coulomb gas of monopoles does not undergo any phase transition between a 
dielectric and metallic phase in three dimensions.  The system is always 
in a ``high-temperature" metallic phase. This can be seen by deriving the 
corresponding recursion relations for a three-dimensional Coulomb gas. The 
recursion relations for a $d$-dimensional Coulomb gas were derived by 
Kosterlitz \cite{Kosterlitz}. The result is

\begin{equation}
\label{flowK}
\frac{d K^{-1}}{dl}=4\pi^2 y^2-(2-d)K^{-1},
\end{equation}

\begin{equation}
\label{flowy}
\frac{dy}{dl}=\left[d-2\pi^2f(d)K\right]y,
\end{equation}
where $f(d)=(d-2)\Gamma[(d-2)/2]/(4\pi)^{d/2}$. 
Here, $y(l)$ and $K(l)$ are essentially the scale-dependent fugacity and 
inverse dielectric constant of the $d$-dimensional Coulomb gas, respectively. 
These renormalization group equations are therefore basically 
nothing but self-consistency equations for scale-dependent electrostatics. For 
$d=2$, the above equations reduce to the celebrated recursion relations for 
the Kosterlitz-Thouless (KT) phase transition \cite{KT}. In this case a line 
of fixed points occurs in the flow diagram. However, for $d > 2$ there are 
no zero-fugacity fixed points to the above recursion relations. For $d > 2$, 
the second term  on the right hand side in Eq. (\ref{flowK}) is positive, 
which means that even at arbitrarily low fugacity, $K^{-1}(l) > 0$ will 
increase indefinitely as the (logarithmic) length scale $l$ increases. Hence, 
no matter how large we make the bare value of $K(l)$, it will eventually be 
reduced enough, by Eq. (\ref{flowK}), to make the right hand side of Eq. 
(\ref{flowy}) positive. Hence, it is inevitable that $y(l)$ will eventually 
start increasing with $l$, thus destroying the zero-fugacity fixed point well 
known to exist in the case $d=2$ \cite{KT}.  In particular, this result means 
that no phase transition occurs in $d=3$ in the ordinary Coulomb gas. 
Thus, the Coulomb gas of monopoles is always in the plasma phase. This 
is nothing but a statement which is equivalent to permanent confinement 
of electric test charges in compact three-dimensional photodynamics 
\cite{Polyakov}.

When matter fields are present, the confinement properties 
of the theory are likely to be changed. There are, however, 
many subtleties involved when the matter fields are in 
the so called fundamental representation of compact $U(1)$. 
To see what are the main points, let us consider the lattice 
abelian Higgs model, whose action is given by

\begin{equation}
\label{Higgs}
S=-\beta\sum_{x,\mu}\cos[\Delta_\mu\theta(x)-qA_\mu(x)]
-\kappa\sum_{x,\mu,\nu}\cos[F_{\mu\nu}(x)],
\end{equation}
where $q\in\mathbb{N}$ is the charge carried by the Higgs field . 
The case where the Higgs field carries the 
fundamental charge ($q=1$) differs 
in an essential way from the case $q>1$. 
This can be seen by considering the limiting cases $\beta\to\infty$ 
and $\kappa\to\infty$. Let us consider first the case $q=1$. First 
of all, for all values of $q$ the limit $\kappa\to\infty$ corresponds 
to the $3DXY$ model. This is so because when $\kappa \to \infty$
all gauge-field fluctuations are supressed except those that 
are indistinguishable from vortex-loop fluctuations in the matter 
sector. Hence, for all $q$, the model exhibits a phase transition 
when $\kappa \to \infty$. The limit $\beta\to\infty$, on the other hand, 
is trivial when $q=1$, and there is no phase transition associated with 
it. 

The situation for $q=2$ is drastically different since in this 
case the limit $\beta\to\infty$ leads to a $Z_2$ gauge theory, 
which exhibits a phase transition in the Ising model universality class 
when the space-time dimensionality $d=3$. Thus, when $q=2$ it is natural 
to think that there is a critical line in the phase diagram of the $q=2$ 
theory that interpolates between the two limiting critical regimes, and 
this can indeed be demonstrated \cite{Bhanot,Sudbo,Smiseth}. The case 
corresponding to the Higgs field with the 
fundamental charge does not have two asymptotic critical regimes 
to be interpolated. For this reason, it is generally thought that there 
is no phase transition in the $q=1$ three-dimensional lattice Abelian Higgs 
model and that therefore the theory permanently confines electric test 
charges \cite{FradShe}, as is the case in the pure compact Maxwell theory 
\cite{Polyakov}.

Another obstacle against a deconfinement phase transition in the $q=1$ 
case comes from Elitzur's theorem \cite{Elitzur}. This theorem 
simply states that averages of non-gauge invariant operators 
are always zero in the absence of gauge fixing. Only gauge-invariant 
operators can have a nonzero expectation value. In other words, 
a local gauge symmetry cannot be spontaneously broken. This is  
important, since it implies that the Higgs mechanism can only occur upon 
some gauge fixing. In the lattice action a natural gauge 
fixing is the unitary gauge parametrization. In the unitary gauge 
there is a residual {\it global} gauge symmetry left, which can in 
turn be broken. In the $q=2$ case the residual global symmetry 
corresponds to the $Z_2$ group. Thus, in this case the Higgs phase 
can be distinguished from the confinement phase \cite{FradShe}. In 
the $q=1$ case, however, the residual symmetry is just the identity 
group and it is therefore trivial. {\it The Higgs phase cannot be 
distinguished from the confinement phase}. Again, there is only one 
phase and it seems that there is no way out from permanent confinement 
for fields carrying the fundamental charge.        

In the case of continuous global symmetries the Mermin-Wagner theorem 
\cite{MW} forbids spontaneous symmetry breaking in two dimensions. 
Elitzur's theorem is far more restrictive than Mermin-Wagner's theorem, 
since it applies to any dimension and to discrete gauge groups. A well 
known way out from the Mermin-Wagner theorem is the KT transition 
\cite{KT}, where a phase transition occurs in the absence of long range 
order. The KT transition occurs precisely in the case of a {\it global} 
$U(1)$ group. 

Since the global symmetries do not suffer the severe restriction 
imposed by Elitzur's theorem, duality transformations where a locally 
gauge invariant theory is mapped into a globally invariant theory is a 
powerful tool. We can look for phase transitions there where the symmetry 
can be spontaneously broken and, in some cases, even to look 
for phase transitions without spontaneous symmetry breaking, 
like the KT phase transition at $d=2$. Unfortunately, 
it seems to be generally the case that the phase transition 
in the dual model corresponds to a nontrivial residual 
symmetry in the original model. Thus, as far as the phase 
transition is concerned, in those cases it can easily be established 
in the original model as well. In such cases, the dual model is still  
a powerful tool to establish the universality class. It would 
be interesting if a KT-like phase transition could occur in the dual 
three-dimensional theory, elevating  the three-dimensional $q=1$ Abelian 
Higgs model to a system sustaining a confinement-deconfinement transition. 
Recently it was pointed out in Ref. \cite{KNS} that such a transition 
indeed occurs in this case. {\it A deconfinement phase transition is 
driven by a KT-like phase transition in the monopole plasma}. 
Note that in general, the topological defects of theory are more 
complicated objects. The monopoles are generally connected 
by magnetic vortex lines, and there are also closed vortex 
loops. However, at the critical point the vortex line looses 
tension and we have once more an effective description in terms of 
a gas of magnetic monopoles. The difference to the usual Coulomb 
gas is that the interaction between the monopoles are no longer 
$\propto 1/r$ as in the pure compact Maxwell theory. The matter 
fields induce an anomalous scaling behavior and the monopole-monopole 
interaction becomes $\propto\ln r$ in three dimensions \cite{KNS,KNS1}. 
This behavior strongly suggests that a KT-like transition may also 
occur for this {\it anomalous Coulomb gas} in three dimensions. 

For low fugacity the anomalous Coulomb gas can be brought in the form 
of a sine-Gordon theory with a $1/|p|^3$ free propagator 
\cite{KNS,KNS1}:

\begin{equation}
\label{ASG}
S=\frac{1}{8\pi^2K}
\int d^3 x[\varphi(-\partial^2)^{3/2}\varphi
-2z\cos\varphi], 
\end{equation}
where $K=1/g^2$, with $g$ being the gauge coupling. From the above sine-Gordon 
theory we could in principle raise the following objection to a KT-like behavior 
at $d=3$. It could happen that a $\varphi(-\partial^2)\varphi$ term is generated 
by fluctuation effects. If the theory behaves in such a way, the generated term 
would clearly dominate at large distances and the $\varphi(-\partial^2)^{3/2}\varphi$ 
would become irrelevant in the renormalization group (RG) sense. The resulting effective 
action would just correspond to an ordinary sine-Gordon theory in three dimensions and, 
therefore, no phase transition occurs in this case. The $\ln r$ interaction is screened 
into a $1/r$ potential. This argument, if correct, would spoil the deconfinement phase 
transition for the $q=1$ case. Let us show that this is not the case by using two 
different arguments. 

The first argument relies on simple 
power counting. It turns out that $d=3$ is the lower 
critical dimension of the problem since the field 
$\varphi$ is dimensionless. This means that renormalization 
proceeds in precisely the same way as in the $d=2$ well known 
counterpart of the present problem. 

The second argument relies 
on {\it exact} scaling and duality properties of the 
theory. The sine-Gordon action (\ref{ASG}) is a result of 
a duality transformation of the {\it critical} effective 
action:

\begin{equation}
\label{anomMaxwell}
S_{\rm eff}\propto\int d^3xF_{\mu\nu}\frac{1}{\sqrt{
-\partial^2}}F_{\mu\nu}.
\end{equation}
The above corresponds to an {\it exact} scaling 
behavior of the gauge field propagator in the three-dimensional 
Abelian Higgs model \cite{Herbut}. Next, we proceed by 
{\it reductio ad absurdum} to prove that the effective 
sine-Gordon action (\ref{ASG}) gives the dominant contribution 
at large distances. This is seen as follows.  
The effective action (\ref{anomMaxwell}) clearly corresponds 
to the dominant large distance behavior, i.e., the usual 
Maxwell term $\propto F_{\mu\nu}^2$ is obviously irrelevant in the 
infrared against the anomalous contribution given in Eq. (\ref{anomMaxwell}). 
Here, it is crucial that the sign of coefficient of the anomalous term is
positive, to ensure that the usual Maxwell term is indeed irrelevant. 
A negative sign in front of the anomalous term would make the usual Maxwell 
term relevant. Next, assume that a  $\varphi(-\partial^2)\varphi$ 
term is generated by fluctuation effects in Eq. (\ref{ASG}) and that  
this term has a positive sign. This clearly implies that the                  
$\varphi(-\partial^2)^{3/2}\varphi$ term becomes irrelevant in the infrared. 
Therefore, the corresponding effective sine-Gordon action is of the usual 
type, {\it thus dualizing back to an ordinary Maxwell theory}. This contradicts 
the exact result that the dominant effective critical theory is given by 
Eq. (\ref{anomMaxwell}).  

Next, to further substantiate our scenario, we derive the corresponding 
recursion relations for a gas of point-charges in three dimensions
interacting through a logarithmic pair-potential, i.e., a three-dimensional
logarithmic plasma or anomalous Coulomb gas, and for which Eq. (\ref{ASG}) 
is a field-theoretical description \cite{KNS}. As in the case of Eqs. 
(\ref{flowK}) and (\ref{flowy}), we will find it useful to consider the problem 
in $d$ dimensions. Furthermore, we will consider a more general propagator 
of the form $1/|p|^\sigma$ for the anomalous sine-Gordon model. Thus, we 
consider a {\it bare} potential given by $U_0(r)=-4\pi^2KV(r)$, where 

\begin{equation}
\label{potential}
V(r)=\frac{\Gamma\left(\frac{d-\sigma}{2}\right)}{
2^\sigma\pi^{d/2}\Gamma(\sigma/2)}[(\Lambda r)^{\sigma-d}-1],
\end{equation}
with $\Lambda$ being 
an ultraviolet cutoff. For the particular case $d=\sigma$, 
corresponding to the lower critical dimension in this generalized problem, we 
have

\begin{equation}
\label{logpot}
V(r)|_{d=\sigma}=-\frac{2^{1-\sigma}\pi^{-\sigma/2}}{\Gamma(\sigma/2)}
\ln(\Lambda r).
\end{equation}
The bare electric field is given by 
$E_0(r)=-4\pi^2K A(d,\sigma)r^{\sigma-d-1}/r_0^{\sigma-d}$, where 
$r_0\equiv 1/\Lambda$ and 
$A(d,\sigma)=(d-\sigma)\Gamma[(d-\sigma)/2]/[2^\sigma\pi^{d/2}\Gamma
(\sigma/2)]$. The  {\it bare} electric field is renormalized by the other 
dipoles which are treated as a dielectric medium. The {\it renormalized} 
electric field is then given by

\begin{equation} 
E(r)=-\frac{4\pi^2K  ~ A(d,\sigma) ~r^{\sigma-d-1}}{\varepsilon(r)}, 
\end{equation}
where $\varepsilon(r)$ is the scale-dependent dielectric constant of the 
medium. We can write this in the form $\varepsilon(r)=1+S_d\chi(r)$, where 
$S_d=2\pi^{d/2}/\Gamma(d/2)$, and the electric susceptibility  

\begin{equation}
\label{suscept}
\chi(r)=N_d\int_{r_0}^r n(s,\theta)\alpha(s) s^{d-1}\sin^{d-2}\theta 
ds d\theta,
\end{equation}
with $N_d=2\pi^{(d-1)/2}/\Gamma[(d-1)/2]$. In Eq. (\ref{suscept}), 
$n(r,\theta)$ is the density of pairs and $\alpha(r)$ is the 
polarizability of a dipole. We have for small separation of the pairs 

\begin{equation}
\alpha(r)=\frac{4\pi^2K r^2}{d}+{\cal O}(r^4).
\end{equation}
For $n(r,\theta)$ we have at lowest order in the bare fugacity $y_0$

\begin{equation}
n(r)=\frac{y_0^2}{r_0^{2d}}e^{-U(r)},
\end{equation}
where $U(r)$ is the effective potential obtained by integrating the 
renormalized electric field:

\begin{equation}
\label{effpot}
U(r)=U(r_0)+4\pi^2
K A(d,\sigma)\int_{r_0}^{r}ds \frac{s^{\sigma-d-1}}{\varepsilon(s)}.
\end{equation}
The effective stiffness $K_{\rm eff}(l)$ is related to the dielectric constant 
$\varepsilon(r)$ through

\begin{equation}
\label{stiffness}
\frac{1}{K_{\rm eff}(l)}=\frac{\varepsilon(r_0\exp l)}{K}e^{-(\sigma-d)l}.
\end{equation}
Let us define $u(l)=U(r_0\exp l)$, out of which we obtain 

\begin{equation}
\label{udel}
u(l)=u(0)+4\pi^2A(d,\sigma)\int_0^l dv K_{\rm eff}(v).
\end{equation}
Thus, 

\begin{equation}
\label{flowu}
\frac{du}{dl}=4\pi^2A(d,\sigma)K_{\rm eff}(l).
\end{equation}
We next define the square of the effective fugacity as follows

\begin{equation}
\label{effecfug}
y^2(l)=\frac{2S_d^2}{d r_0^{\sigma-2}}y_0^2 e^{(2d-\sigma+2)l-u(l)}.
\end{equation}
From Eqs. (\ref{stiffness}), (\ref{flowu}), and (\ref{effecfug})  
we finally obtain

\begin{equation}
\label{flowK2}
\frac{d K_{\rm eff}^{-1}}{dl}=4\pi^2y^2-(\sigma-d)K_{\rm eff}^{-1},
\end{equation}

\begin{equation}
\label{flowy2}
\frac{dy}{dl}=\left[d-\eta_y-2\pi^2A(d,\sigma)K_{\rm eff}\right]y,
\end{equation}
where the {\it anomalous dimension} of the fugacity is given by 
$\eta_y=(\sigma-2)/2$. When $\sigma=2$, we recover the recursion relations 
(\ref{flowK}) and (\ref{flowy}), which were originally derived by a 
completely different method \cite{Kosterlitz}. Such a theory does not 
exhibit a phase-transition for $d>2$. The case relevant for the three-dimensional 
Abelian Higgs model is, on the other hand, $\sigma=d=3$. In this case 
Eqs. (\ref{flowK2}) and (\ref{flowy2}) are very similar to the usual KT 
recursion relations, except for the presence of the anomalous scaling 
dimension of the fugacity, $\eta_y$ which is nonzero in our case and 
given by $\eta_y=1/2$. 

By integration of the above recursion relations Eqs. (\ref{flowK2}), (\ref{flowy2}), 
for the case $d=\sigma=3$, we may compute explicitly the screened effective potential 
$u(l)$ on the  ordered side, the result is \cite{Note}
\begin{eqnarray}
u(l)-u(0) =   
\frac{1}{\omega_- \omega_+} \left[
\frac{5}{2} \omega_+ l + \ln \left(
\frac{\omega_+ e^{-2 \theta} + \omega_-}{\omega_+ e^{-2 u } + \omega_-} \right) \right]
\end{eqnarray}
with $\omega_{\pm} = 1 \pm \omega$, $u = (5/2) \omega l + \theta$, and $\omega$ and $\theta$ 
are integration constants determined from initial conditions on the flow equations. 
Asymptotically, we have $u(l) \sim l \sim \ln(r/r_0)$, which shows that the effective
potential is also logarithmic after screening effects are taken into account. Hence, 
we conclude that the statement, alluded to above, that the transition is destroyed 
by screening a bare $\ln(r)$-potential into a $1/r$-potential, 
is not correct. This can also be seen from a simple Debye-H\"uckel 
theory for a $\ln(r)$-potential in $d$ dimensions. Indeed, due to 
Gauss's theorem 
in $d$-dimensions, the field equation in the corresponding 
Debye-H\"uckel theory is given by \cite{Dixit} 
$\nabla\cdot({\bf E} r^{2-d})=S_d[q\delta^d({\bf r})+
\langle\rho({\bf r})\rangle]$, where $S_d=2\pi^{d/2}/\Gamma(d/2)$ 
and $\langle\rho({\bf r})\rangle$ is the variation of the charge 
density in a plasma with density $n_0$ \cite{Ichimaru}. 
The electric field is 
${\bf E}=-\nabla U$, where $U$ is the screened effective potential. 
The high temperature limit corresponds to the Debye-H\"uckel 
approximation, where the differential equation for $U(r)$ 
can be solved exactly \cite{Dixit}. The result is 

\begin{equation}
\label{DHpot}
U(r)=\frac{2q}{d}K_0[(r/\lambda_D)^{d/2}]-q\ln\lambda_D,
\end{equation}
where $K_0$ is a modified Bessel function of the second kind and
 the inverse of the Debye screening length  $\lambda_D$ is given by
\begin{eqnarray}
\lambda_D^{-1}=\left(\frac{16q^2n_0\pi^{d/2}}{d^2 ~T \Gamma(d/2)} \right)^{1/d},
\end{eqnarray} 
where $T$ is the temperature. For $r/\lambda_D\ll 1$, 
Eq. (\ref{DHpot}) has the expansion 
$U(r)=2q(\ln 2-\gamma)/d-q\ln r+{\cal O}(r^2/\lambda_D^2)$.    

The recursion relations Eqs. (\ref{flowK2}) and (\ref{flowy2}) thus predict the 
possibility of a topological phase-transition in the system of point charges 
interacting with the potential Eq. (\ref{potential}) from a ``low-temperature" 
dielectric phase to a ``high-temperature" metallic phase, with a universal jump in the stiffness 
of the theory given by  $K^*_{\rm{eff}} = 2/5$. This should be contrasted with
the universal jump $K^*_{\rm{eff}}=2/\pi$ that is found in the two-dimensional
case. A specific realization of 
such a KT-like scenario has recently been suggested in the context of 
discussing the physics of strongly correlated fermion systems at zero
temperature in two spatial dimensions \cite{KNS}.

In summary, we have proposed a mechanism in three 
space-time dimensions by which a deconfinement 
transition occurs for $U(1)$ compact gauge fields coupled to 
bosonic fundamental matter fields. The proposed mechanism 
relies on a KT-like phase transition in three space-time dimensions. 
Therefore, no symmetry breaking is involved and Elitzur's theorem 
is not violated. Instead, a topological phase transition occurs.

\end{document}